\documentclass[manuscript]{aastex61}
 \usepackage{graphicx, subfigure}
 \usepackage{natbib}
 \usepackage{times}
 \usepackage{amsmath}
 \usepackage{textcomp}

\submitjournal{ApJ}
\shorttitle{ New  Photo-zs with $Swift$ and SARA}
\shortauthors{Kaur et al.}

\def\asec{\ifmmode^{\prime\prime}\else$^{\prime\prime}$\fi}
\def\degs{\ifmmode ^{\circ}\else$^{\circ}$\fi}

\begin{document}

\title{New high-$z$ BL Lacs using the photometric method with $Swift$ and SARA}

\author{A. Kaur}, 
\affil{Department of Physics and Astronomy, Clemson University, SC 29634-0978, U.S.A.}
\email{akaur@g.clemson.edu} 
\author{ A. Rau},
\affil{Max-Planck-Institut f\"ur extraterrestrische Physik, Giessenbachstra{\ss}e 1, 85748 Garching, Germany}
\author{M. Ajello},
\affil{Department of Physics and Astronomy, Clemson University, SC 29634-0978, U.S.A.}
\author{ A. Dom\'{i}nguez},
\affil{Grupo de Altas Energ\'{i}as, Universidad Complutense, E-28040 Madrid, Spain}
\author{V. S. Paliya}, 
\affil{Department of Physics and Astronomy, Clemson University, SC 29634-0978, U.S.A.}
\author{J. Greiner},
\affil{Max-Planck-Institut f\"ur extraterrestrische Physik, Giessenbachstra{\ss}e 1, 85748 Garching, Germany}
\author{D. H. Hartmann},
\affil{Department of Physics and Astronomy, Clemson University, SC 29634-0978, U.S.A.}
\author{ P. Schady}
\affil{Max-Planck-Institut f\"ur extraterrestrische Physik, Giessenbachstra{\ss}e 1, 85748 Garching, Germany}
 
\begin{abstract}
BL Lacertae (BL Lac) objects are the prominent members of the third {\it Fermi}
Large Area Telescope catalog of $\gamma$-ray sources. Half of the BL
Lac population ($\sim$ 300) lack redshift measurements, which is due
to the absence of lines in their optical spectrum, thereby making it
difficult to utilize spectroscopic methods. 
Our photometric drop-out technique can be used to establish the
redshift for a fraction of these sources.
This work employed 6 filters mounted on the $\emph{Swift}$-UVOT and 4
optical filters on two telescopes, the 0.65 m SARA-CTIO in Chile and 1.0
m SARA-ORM in the Canary Islands, Spain. A sample of 15 sources was
extracted from the $\emph{Swift}$ archival data for which 6 filter
UVOT observations were conducted. By complementing the {\it Swift}
observations with the SARA ones,  we  were able to discover two high
redshift sources: 3FGL J1155.4-3417 and 3FGL J1156.7-2250 at 
$z=1.83^{+0.10}_{-0.13}$ and $z=1.73^{+0.11}_{-0.19}$ ,
respectively, resulting from the dropouts in the powerlaw template fits to these data. The discoveries add to the important (26 total) sample of
high-redshift BL Lacs. While the sample of high-z BL Lacs is still rather small, these objects do not seem to fit well within known schemes of the blazar population and represent the best probes of the extragalactic background light.
\end{abstract}
\section{Introduction} \label{sec:intro}
Active Galactic Nuclei (AGN) possessing jets aligned with our line of
sight are known as blazars \citep{Blandford1978}. The spectral energy
distribution of these sources displays two broad bumps attributed to
synchrotron emission at low energies (IR to X-ray) and the inverse
Compton scattering at high energies \citep[X-ray to $\gamma$-rays][]{Abdo2011, Abdo2011a}.   \citet{Abdo2010} introduced a classification criteria for 
blazars based on their peak synchrotron frequencies,
$\nu_{pk}^{sy}$. These authors divided blazars into three classes;
high-synchrotron-peak (HSP, $\nu_{pk}^{sy} > 10^{15}$ Hz),
intermediate-synchrotron-peak, (ISP, $10^{14} < \nu_{pk}^{sy} <
10^{15}$ Hz) and low-synchrotron-peak (LSP, $\nu_{pk}^{sy} < 10^{14}$)
Hz objects. Another classification scheme for blazars divides them
into two classes due to the differences in their optical spectrum
properties: BL Lacertae objects and Flat Spectrum Radio Quasars
(FSRQs). The former exhibit no or very weak ($<$5 \AA equivalent width) emission
lines \citep{Urry1995}, whereas the latter show broad emission
lines. The absence of lines in BL Lacs implies either their spectrum
is dominated by the synchrotron emission from the jet \citep{Marcha1996}
or that the emission from the disk and broad line region is very weak due to,
likely, inefficient (or low) accretion, jet dilution or a combination of the the both scenarios \citep{Giommi2011}. Most of the
FSRQs are identified as LSPs, whereas BL Lacs can belong to all three classes (LSP, ISP and HSP) with most BL Lacs exhibiting ISP and HSP characteristics, i.e. substantial emission at $>$ 10 GeV \citep{Ackermann2015b}. This property makes BL Lacs bright $\gamma$-ray emitters and hence excellent probes of the extragalactic background light (EBL \citet{Dominguez2015}), which consists of all the emission from the stars and accreting objects in the observable universe since galaxy formation. \\
The direct measurement of EBL is challenging due to the bright
zodiacal light and emission from our galaxy \citep{Hauser2001}. An indirect method
to study the EBL intensity is by using $\gamma$-ray emitters. The underlying principle for this method
utilizes the interaction of the $\gamma$-ray photons interact with the EBL
photons to produce electron-positron pairs. This leads to a characteristic attenuation in the $\gamma$-ray spectra \citep{Stecker1992, Ackermann2012}. This
imprint can be used to study the EBL and its evolution with redshift
\citep{Aharonian2006}. Moreover, sources at higher redshifts are more
strongly attenuated, thereby leading to better EBL
constraints. High-redshift BL Lacs, because of their prominent
emission at $>$10\,GeV are thus the best probes of the EBL, but they
are rare.
So far only 24 BL Lacs, among the 700 {\it Fermi}-LAT 3LAC sources \citep{Ackermann2015b},
have $z>1.3$. Several authors utilized various facilities to obtain the redshifts for these objects using the spectroscopic method, but only telescopes greater than 8m class yielded redshifts measurements. Otherwise lower limits were placed due to the faint absorption lines from the galaxies along our line of sight, \citep[e.g.][s]{Landt2012,Shaw2013a,Shaw2013b,Massaro2015,Crespo2016a}.  \\
 The photometric method for calculating redshifts for BL Lacs was
 first introduced by \citet{Rau2012}, which used quasi-simultaneous observations with UVOT \citep[Ultaviolet and Optical Telescope][]{ Roming2005} mounted on the Neil Gehrels Swift Observatory \citep{Gehrels2004}
 and GROND \citep{Greiner2008}, a multicolor imager at the 2.2m
 MPG telescope, mounted at the ESO La Silla Observatory, Chile. This inexpensive and
 time efficient method is based on the following principle: the UV
 photons from BL Lacs are absorbed by the neutral hydrogen along our
 line of sight, thereby attenuating the flux bluewards of the Lyman
 limit. This dropout in the spectral energy distribution of a BL Lac
 can be modeled to obtain a (photometric) redshift. These
 authors found 9 (6 new) high redshift BL Lacs (out of 103), increasing the total
 sample of high-$z$ BL Lacs by 50\% utilizing $\sim$ 1ks
 $\emph{Swift}$-UVOT exposure time per source for all  filters
 combined and 4-5 min with GROND. Upper limits (typically $z<1.2-1.3)$ were established for the rest of the sources in that work. \citet{Kaur2017} continued this work and found 5 more high-$z$ BL Lacs from a sample of 40 objects. 
 
The work presented here is the continuation of this successful
program. In this new analysis, we rely for the first time on archival $\emph{Swift}$-UVOT data and two ground based facilities, as explained in the next section. This paper is organized as follows: section 2 introduces the facilities, data collection and the observing strategy. Section 3 explains the data analysis steps used for the completion of this work. Section 4 illustrates the SED fitting, followed by the results in Section 5. The paper is concluded in the Section 6 with a brief discussion of our findings. It should be noted that a flat $\Lambda$CDM cosmological model with H$_0$=0.73 km s$^{-1}$ Mpc$^{-1}$, $\Omega_{m}$=0.27, $\Omega_{\Lambda}$=0.73 was adopted for all the calculations in this work.

\section{Sample Selection} \label{sec:sample}
 First, the HEASARC archive was explored to search for blazars,
 classified as BL Lacs in the 3FGL catalog \citep{Acero2015}. Sources
 observed in all 6 (\emph{uvw2, uvm2, uvw1, u, b, v}) filters of UVOT mounted on the Neil Gehrels $\emph{Swift}$ observatory were selected and not included in \citet{Rau2012} or \citet{Kaur2017}. These BL Lacs were then observed using two ground facilities: the Southeastern Association for Research in Astronomy consortium's  0.65 m and 1.0 m telescopes at Cerro Tololo, Chile (SARA-CT) and Roque de los Muchachos Observatory, Canary Islands (SARA-ORM), respectively. The data were obtained in 4 SDSS filters ( {\it $g^\prime,  r^\prime,  i^\prime, z^\prime$}) mounted on these two telescopes. See \citet{Keel2016} for further details on these two ground facilities. 
 
 A sample of 15 BL Lacs was selected from $\emph{Swift}$ archive based on the above mentioned criteria. These objects were then observed with the two ground based facilities, with exposures times ranging  from 20--60 minutes per filter.  \\
 The details of observations for all the facilties are presented in Table ~\ref{tab:data}. The combined data from the $\emph{Swift}$ satellite and SARA telescopes resulted in a 10 filter flux measurements for each object. 
 
 \section{Method}\label{sec:method}
 \subsection{$\emph{Swift}$-UVOT}
 The standard UVOT pipeline procedure \citep{Poole2007} was followed to extract the final products, which were flat fielded and corrected for the system response. The magnitudes were derived using the UVOT task, {\tt\string UVOTMAGHIST} from \software{HEASoft v.6.21\footnote{https://heasarc.nasa.gov/lheasoft/}} using a circular aperture of variable radius for each object to maximize the signal to noise ratio. The background subtraction was performed by selecting an annular region with inner and outer radius 10'' and 25'' for each source. A careful analysis was performed to select the background aperture to avoid any contamination from the nearby objects in the field. These extracted magnitudes were corrected for the Galactic foreground extinction using table 5 presented in \citet{Kataoka2008} and then converted to the AB system, reported in Table ~\ref{tab:mags}.
 \subsection{SARA-ORM and SARA-CT}
 The data from the two ground facilities were analyzed using the standard aperture photometry technique employing \software{IRAF (v2.16; \citep{Tody1986,Tody1993}}. Photometric calibrations were performed for each object using 1-2 standard stars observed each night. The Sloan filters (g, r, i, z) mounted on the two telescopes were employed for the measurements. 
 The foreground galactic extinction was applied using the calculations from \citet{Schlafly2011}. The final magnitudes in the AB system are shown in Table ~\ref{tab:mags}.
 \subsection{Variability correction}
 In our previous work, \citet{Rau2012} and \citet{Kaur2017}, a special
 measure was taken to observe each source in all the filters
 simultaneously from the ground based facility and within  1-10 days
 from the $\emph{Swift}$ observations. In this work, we are utilizing
 archival $\emph{Swift}$-UVOT  data for the BL Lacs and the
 observations using ground telescopes were performed within the last
 one year. In order to account for the uncertainties due to the
 variable nature of blazars, we applied an additional systematic uncertainty of
 $\Delta$m = 0.1 mag for each UVOT filter, which was established by
 \citet{Rau2012}. 

 Moreover, our previous works utilized the overlap between the
 g$^\prime$ and b$^\prime$ filters in GROND \citep{Greiner2008} and
 $\emph{Swift}$-UVOT for the inter-calibration between the two
 instruments \cite[see][]{Kruhler2011}. Since the GROND and SDSS filters yield the same measurements in g$^\prime$, r$^\prime$, i$^\prime$, z$^\prime$ filters, we employ the same relationship as shown in equation \ref{eqn:1} to calibrate $\emph{Swift}$-UVOT data with respect to the SARA observations. The offsets calculated from this equation in the b band were applied to the all the UVOT filters. The resulting AB magnitudes from both instruments are provided in Table \ref{tab:mags}.
\begin{equation}\label{eqn:1}
b-g^\prime = 0.15\,(g^\prime-r^\prime)+0.03\,(g^\prime-r^\prime)^2  
\end{equation}

\begin{figure}[h!]
	
	\includegraphics[width=0.9\columnwidth]{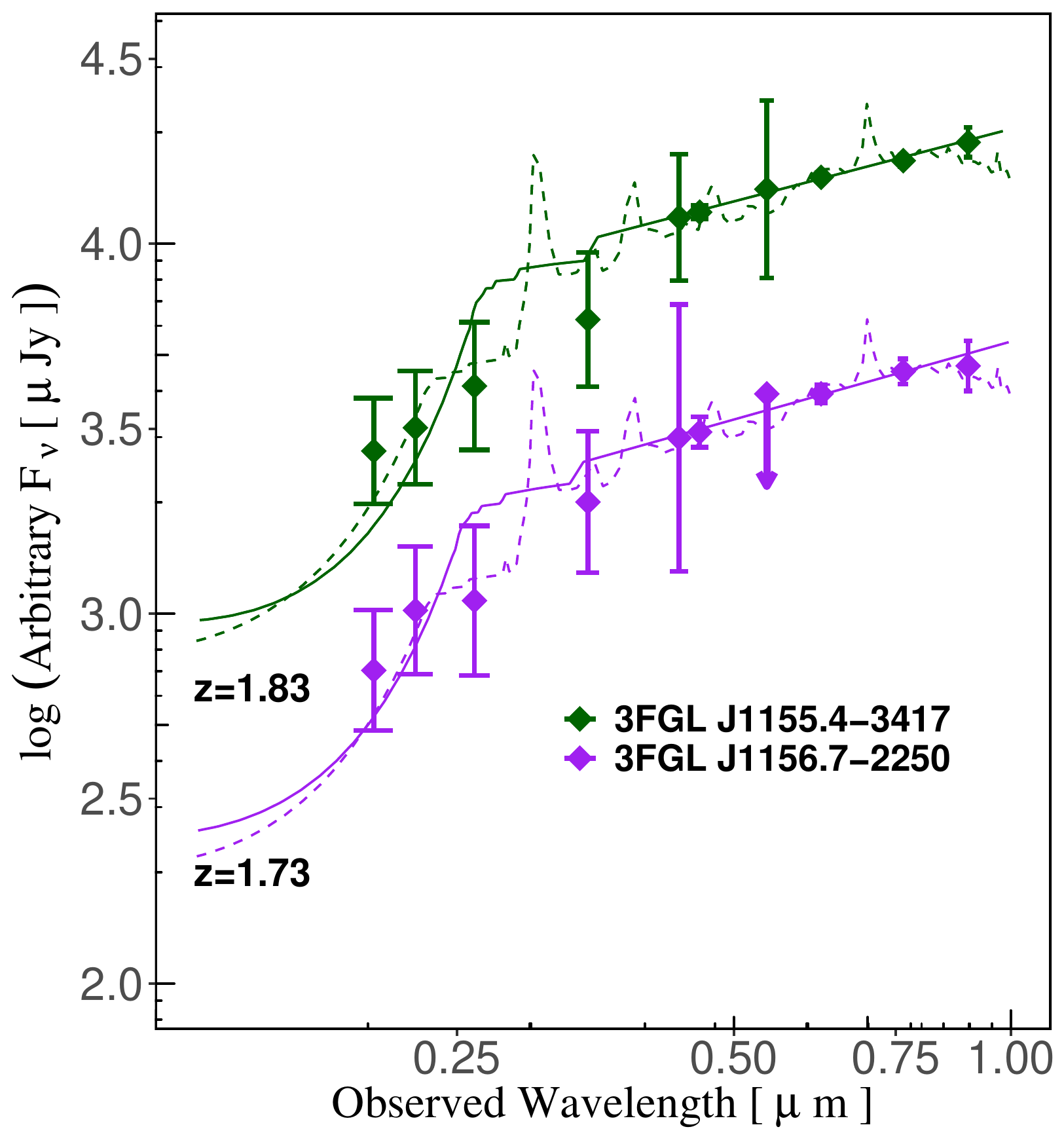}
		\caption{\label{fig:sed} The  $\emph{Swift}$-UVOT +SARA spectral energy distribution of the two  high-$z$ sources. The ($solid~line$) power law templates are fitted for each source with a clearly seen dropout towards the shorter wavelengths. In addition, the ($dashed~line$) galaxy template fits to these objects are presented. The redshifts estimates provided by the latter fits suggest $z=1.5$ for each source. The details of the fitting method and the inference of these results are described in Section ~\ref{sec:sedfit}. and ~\ref{sec:result}., respectively.}
   
\end{figure}
\section{SED fitting}\label{sec:sedfit}
Under the assumption that the spectral energy distribution (SED) does
not change with flux changes from UV-Opt-nearIR regime for the BL Lacs and that it is a
dominated by the non-thermal synchrotron emission, the 10 filter data
obtained in this work could be assumed to follow a power-law spectrum
(typical of BL Lacs).
The fitting program, \software{LePhare v.2.2; \citep{Arnouts1999,Ilbert2006}} was utilized to determine the photometric redshifts. This program is based on the $\chi^2$ statistics for evaluating the difference between the observational and theoretical models. It should be noted that this program includes the response curves for all the filters. Therefore, the red leaks  for UVW1 and UVW2 are taken into account during the fitting procedure. In the context of this work, we employed three different libraries to fit our data. The first library consisted of 60 power-law templates of the form $F_{\lambda} \propto \lambda^{-\beta}$. The value of $\beta$ was chosen to be in the range 0 to 3, since the typical indices for BL Lacs in this wavelength regime are in the above mentioned range. The dropout from the typical power-law fitting was employed to estimate the redshifts. The second and the third libraries comprised of galaxy and stellar templates, respectively, were also fit to check any false associations. The galaxy templates were derived from \citet{Salvato2009,Salvato2011} and the stellar templates were obtained from \citet{Pickles1998,Bohlin1995} and \citet{Chabrier2000}. \\
\citet{Rau2012} performed Monte Carlo simulations for 27000 test SEDs with $\beta$ between 0.5-2.0 and redshifts from 0 to 4, in order to test the reliability of this fitting procedure to determine the photometric redshifts. These simulations yielded that the sources with redshifts greater than 1.2 reproduced the results within an accuracy of$\left| \Delta z(1+z_{sim}) \right| < 0.15 $. Another selection criteria for measuring the reliability of redshift estimates were based on the integral of the probability distribution function,  P$_{z} = \int f(z) dz$ at $z_{phot} \pm 0.1(1+z_{phot})$. The resulting values with $P_z > 90 \%$ were considered reliable for this work.

\begin{figure}[t!]

	\includegraphics[width=\columnwidth]{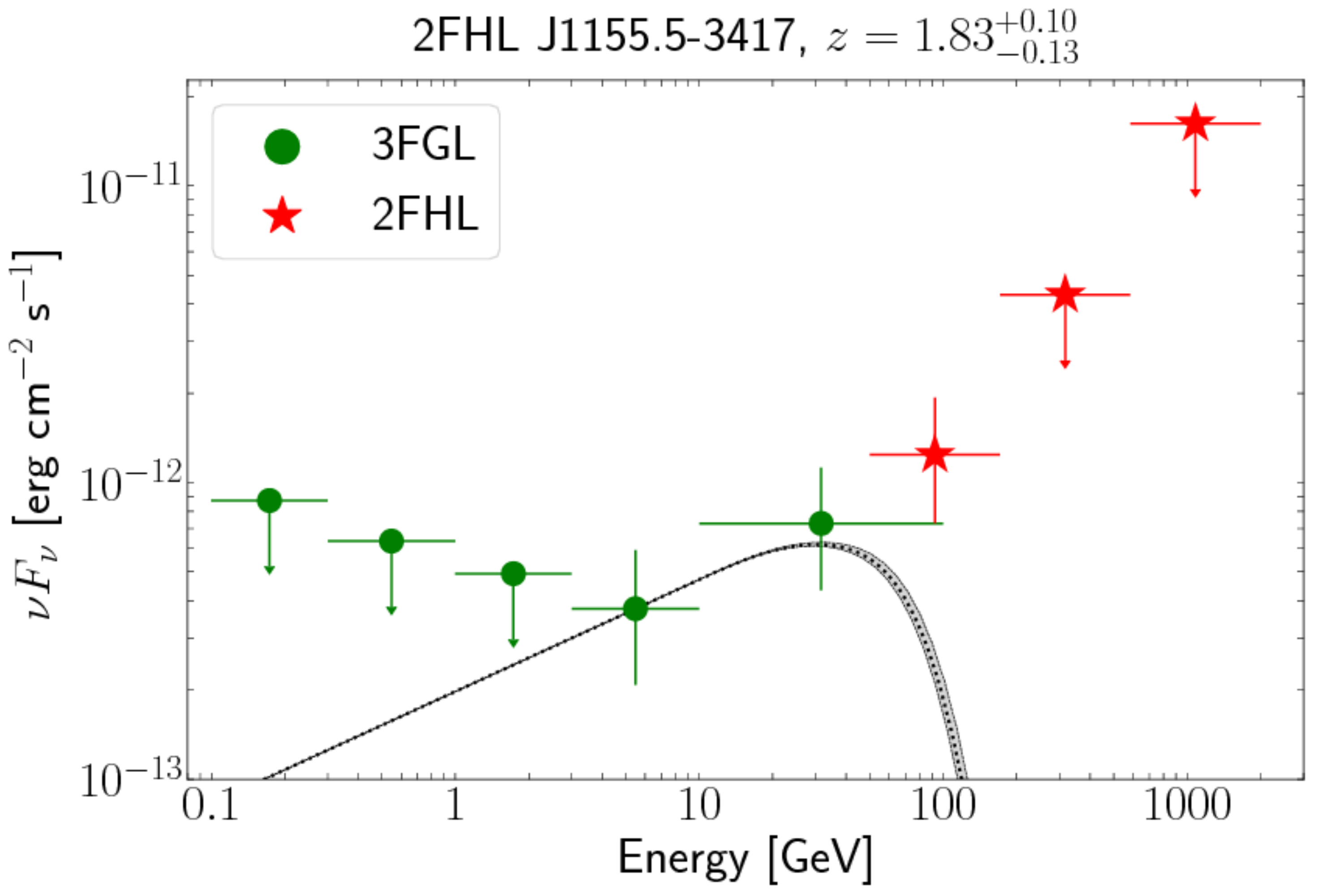}
			\caption{\label{fig:1155}Spectral energy distribution of 3FGL J1155.4-3417 using the 3FGL and 3FHL catalogs. The black line is a power law fit to the 3FHL data, which is absorbed by the EBL, utilizing \citet{Dominguez2011}.}
\end{figure}
\begin{figure}
	\includegraphics[width=\columnwidth]{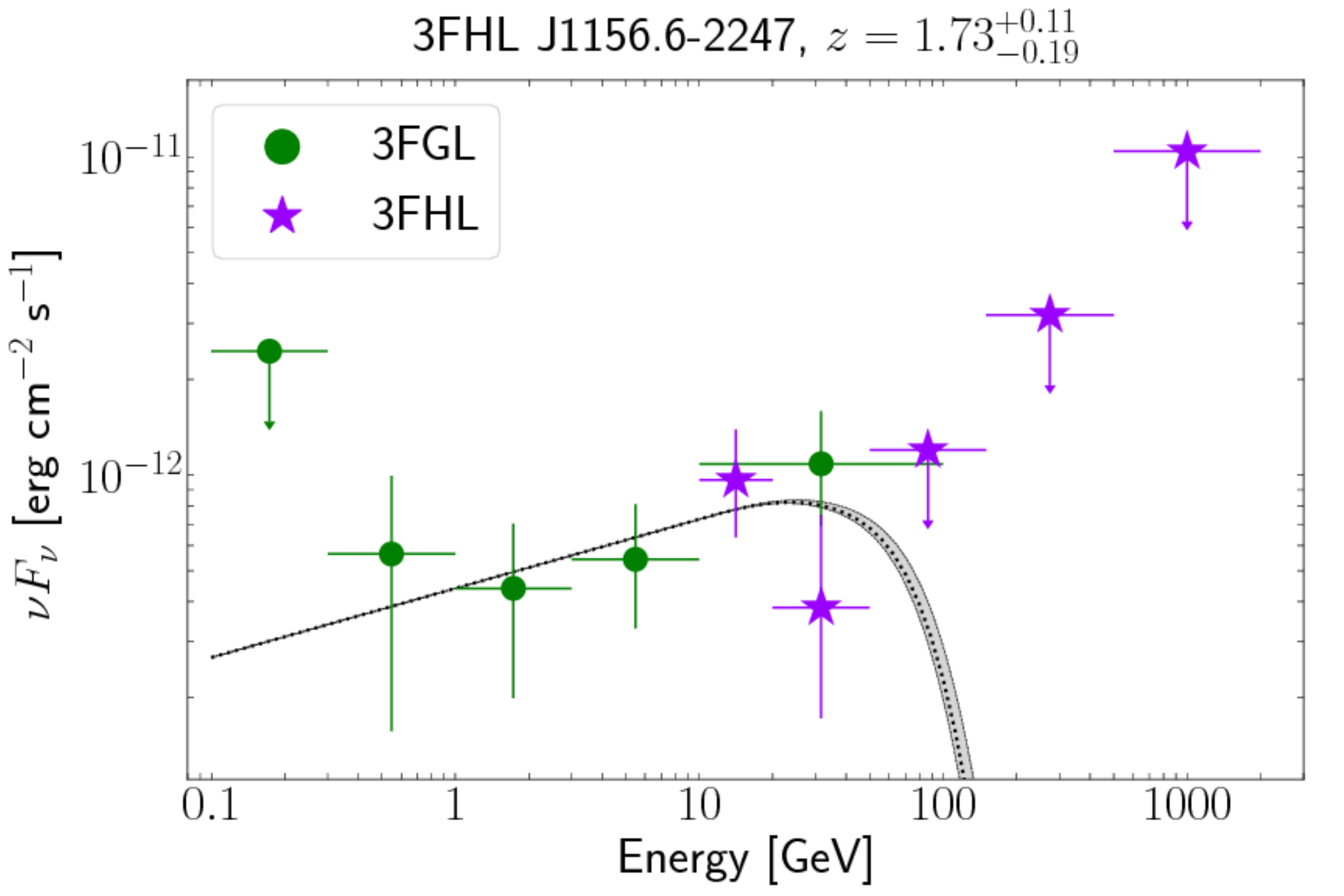}
	\caption{\label{fig:1156}Spectral energy distribution of 3FGL J1156.6-2247 using the 3FGL and 3FHL catalogs. The black line is a power law fit to the 3FHL data, which is absorbed by the EBL, utilizing \citet{Dominguez2011}.}
	
\end{figure}

\section{Results}\label{sec:result}
The SED fitting results were assessed for reliability based on the two
criteria mentioned in the previous section, i.e. $z \geq 1.3$ and $P_z
> 90\%$. From our sample of 15 sources, we found two at redshifts
greater than 1.3, which is consistent with our predicted rate of
10-15\% of the sources being at high redshifts, as indicated by our
previous studies \citep{Rau2012} and \citep{Kaur2017}.  $\emph{Swift}$-UVOT +
SARA SEDs of these two objects,  3FGL J1155.5$+$3417 and 3FGL
J1156.6$-$2247, are shown in Fig.~\ref{fig:sed}. The redshifts for
these two BL Lacs were found to be $z=1.83^{+0.10}_{-0.13}$ and $z=1.73^{+0.11}_{-0.19}$, respectively, using the powerlaw template fits.  In addition, galaxy templates were fit to these data, which yielded redshifts estimates, $z=1.50^{+0.02}_{-0.03}$ and $z=1.50^{+0.04}_{-0.07}$ assuming  hybrid QSO templates from \citet{Salvato2011}. Both the sources, 3FGL J1155.5-3417 and 3FGL J1156.6-2250 are identified as BL Lac \citep{DAbrusco2014} and HSP BCU II \citep{Ackermann2015b} (associated with the BL Lac class), respectively, therefore the redshifts estimates provided by the QSO templates with prominent broad emission lines are rather unlikely. Moreover, the QSO templates yield redshifts estimates without including any fit to the optical emission lines (See Fig.~\ref{fig:sed}), we therefore assume the redshift estimated provided by the powerlaw fits to be more precise. Therefore the rest of the analysis will be based on the redshift estimates derived from powerlaw fits . 
Moreover, upper limits for 4 sources were established. All these results are presented in Table~\ref{tab:result}. None of the sources in our sample were consistent with the stellar templates used for the SED fitting, therefore those results are not displayed in Table~\ref{tab:result}. \\
     The outcome of the combined analysis of BL Lac observations taken at different epochs can be affected by spectral variability. To verify our findings of two new high-z sources, we obtained new observations for these two sources using $Swift$ and SARA simultaneously on  2017-12-28 and 2017-12-27, respectively. The results of this analysis yielded $z=1.73^{+0.11}_{-0.15}$ and $z=1.83^{+0.12}_{-0.15}$ for 3FGL J1155.5$+$3417 and 3FGL
J1156.6$-$2247, respectively, which are consistent with our archival data analysis, within the uncertainties.
\begin{figure}[h]
 	\includegraphics[width=\columnwidth]{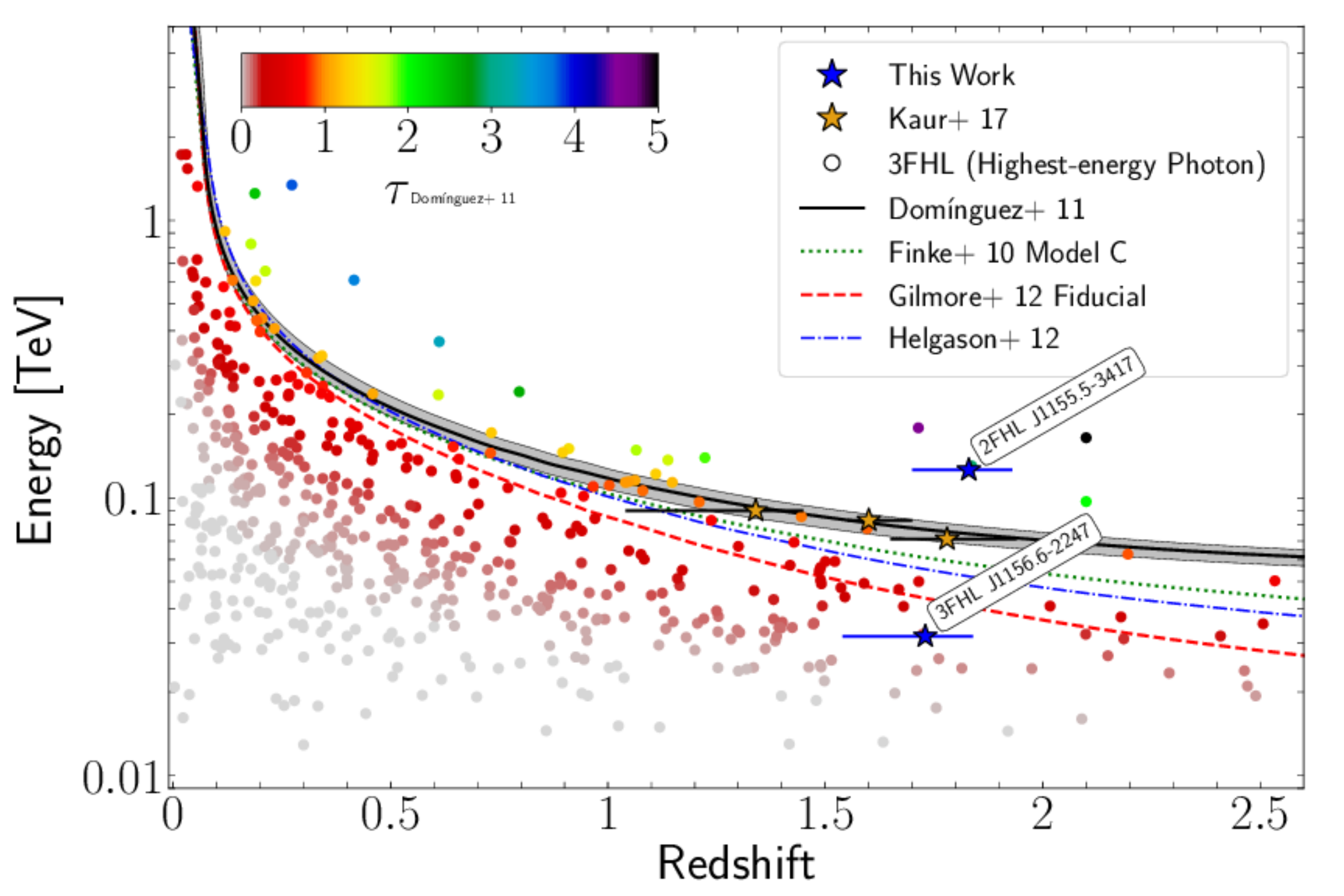}
	 \caption{\label{fig:cgrh}The Cosmic $\gamma$-ray horizon plot: The highest energy of photons from sources (with E $>$ 50\,GeV) vs. their redshift . The colors of sources imply their corresponding optical depth ($\tau$) values (see colorbar). Various estimates of the cosmic $\gamma$-ray horizon, obtained from the EBL models by  \citet[dotted green line]{Finke2010}, \citet[solid black line, with uncertainties as shaded band]{Dominguez2011}, \citet[dashed red line]{Gilmore2012} and \citet[dot-dashed blue line]{Helgason2012} are plotted for comparison. The highest energy photons from one of the two high-$z$ sources lie above the cosmic $\gamma$-ray horizon ($blue~ filled~stars$), whereas the other source lies below this limit. The three $orange$ star symbols represent the high-$z$ BL Lacs found in \citet{Rau2012} and \citet{Kaur2017}, for which 2FHL data were available.}
	 
\end{figure}
\begin{figure}[t!]
	\includegraphics[width=\columnwidth]{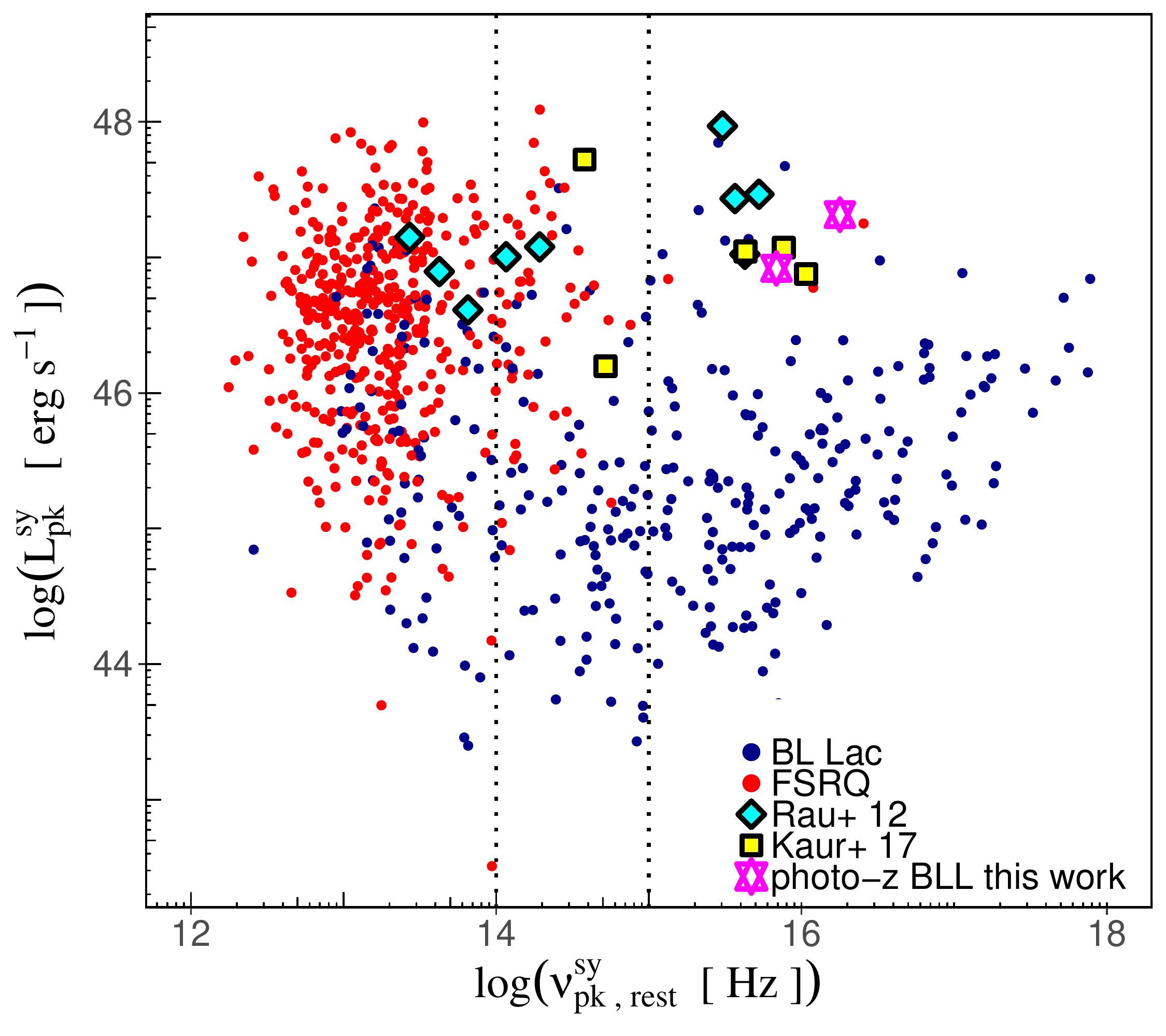}
	\caption{\label{fig:vLv} The peak synchrotron frequency in the rest frame ($\nu_{sy}^{pk}$) vs the peak synchrotron luminosity ($L_{sy}^{pk}$). The magenta stars represent the two new sources found in the present work. These are consistent with the other high-$z$ BL Lacs using photometric method, which display the high luminosity and high synchrotron peak behavior. We show 6 new BL Lacs from \citet[][cyan diamonds]{Rau2012}, 5 BL Lacs from \citet[][yellow squares]{Kaur2017}, and all the FSRQs and BL Lacs from the 3LAC catalog with known redshifts (red and blue circles, respectively). The separation of the LSP, ISP, and HSP regions are also plotted (dotted lines).}

\end{figure}

\begin{figure}[h!]
	
	\includegraphics[width=\columnwidth]{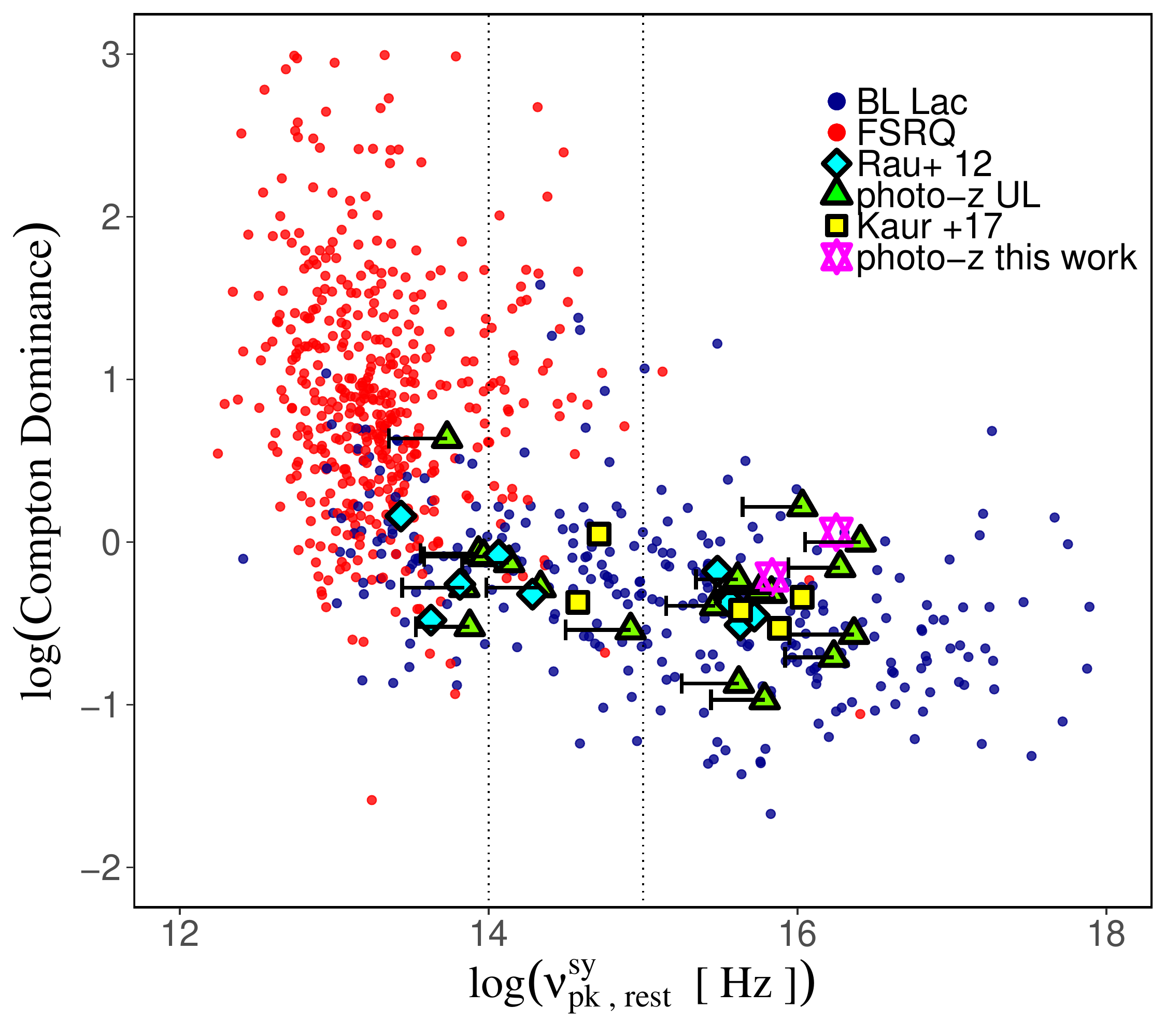}
	\caption{\label{fig:CD} The relationship between the Compton Dominance and  $\nu_{sy}^{\rm pk}$. The color scheme for data symbols follows from Figure~\ref{fig:vLv}. All the upper limits, including the ones from this work are represented in green $ filled$ triangles. The black $horizontal$ lines represent the range of  $\nu_{sy}^{pk}$ for redshifts from zero to the upper limits, provided in our sample.}
\end{figure}

\section{Discussion and Conclusions}
The two newly discovered high-$z$ BL Lacs are also part of the recent
third catalog of high-energy sources
\citep[3FHL,][ 3FGL J1155.5+3417 is also present
in the 2FHL \citet{Ackermann2016a} catalog]{Ajello2017}. Therefore, we utilized the 3FGL and 3FHL
catalogs to construct the SEDs of these two sources. These were fitted
with a power law with an EBL absorption of form, $e^{-\tau(E,z)}$
using the \citet{Dominguez2011} model as illustrated in
Fig.~\ref{fig:1155} and \ref{fig:1156}. These fits indicates that the observed flux at high energies from these two sources was reduced by a factor of $\sim$ 10, due to the EBL absorption, in both  cases. We also reproduced the cosmic gamma ray horizon plot (i.e. the
redshift at which, for a given energy, the Universe becomes opaque to
$\gamma$ rays, see \citet{Dominguez2013} using all the sources in the 3FHL catalog and including
the two new sources found here. This is displayed in Fig.~\ref{fig:cgrh}, which shows that the photometric method is able
to find $\gamma$-ray sources that can constrain the cosmic $\gamma$-ray 
horizon (CGRH), which is defined by the energy at which the optical depth $\tau$ is one as a function of redshift, at redshifts where data are scarce. Interestingly, the highest energy photon from J1155.4$-$3417 lies above the CGRH with $\tau\approx 2.8$. The objects discovered with our photometric technique \citep[the two reported here and also those in][]{Rau2012,Kaur2017} allow us to probe a region of the cosmic $\gamma$-ray horizon where measurements are scarce and as such they allow us to better constrain the EBL. Indeed, the detection of many high-energy photons above the horizon may imply that the EBL model used may be too opaque to characterize the real level of the EBL in the Universe. In the case of Fig.~4, the model of \citet{Gilmore2012} is less favored (because more opaque) than  the one of \citet{Dominguez2011}.
To understand how these high-$z$ BL Lacs fit within the larger blazar
population \citep{Maraschi1995, Sambruna1996,Fossati1998},  we calculated various parameters like the synchrotron
peak frequency $\nu_{sy}^{pk}$, the luminosity at the synchrotron peak
($L_{sy}^{pk}$) and the Compton Dominance (CD, the ratio between the
luminosities of the inverse Compton and synchrotron
peaks). \citet{Fossati1998} observed some correlations between the
above mentioned parameters using the available blazar data. This
sequence indicates that the more luminous blazars possess lower
synchrotron frequencies, but substantial $\gamma$-ray emission (FSRQs
in general), whereas BL Lacs are less luminous, but achieve larger
synchrotron peak frequencies. We test these correlations by plotting
all the blazars in the 3LAC catalog, together with the high-$z$ BL
Lacs discovered in all our photometric campaigns. Fig.~\ref{fig:vLv}
suggests the presence of large luminosity-high frequency synchrotron peak blazars, which do not fit very well with the known blazar sequence scheme. Although, in order to draw a more robust conclusion in this scenario, a larger sample of such sources is required.  Fig.~\ref{fig:CD} displays peak synchrotron frequency vs compton dominance and shows that all our BL Lacs display CD $\leq$ 1, which is consistent with the typical SED of this subclass of blazars, suggesting a more dominant synchrotron emission.  \\
Including the results of this work, our continuing photometric method has discovered 13 BL Lacs.  Overall, 26 (including 2 found here) BL Lacs are known in the literature with $z > 1.3$, of which 50\% were provided by us. Moreover, the accuracy of this method was shown in Fig.2 in \citet{Kaur2017} by successfully matching the spectroscopic and photometric redshifts estimates for the BL Lacs.
  
In the present 3LAC catalog \citep{Ackermann2013a}, about 300 BL Lacs lack redshift measurements.  The primary objective of our photometric program is to provide the redshifts or at least the upper limits for the 3LAC catalog for all the BL Lacs. Based on the previous results, this work has yielded a total of 16 sources (9 (6 new) from \citet{Rau2012}, 5 from \citet{Kaur2017}, 2 from this work) with high redshifts from sample sizes of 103, 40 and 15, respectively. Based on the probability of finding 10-15\% of BL Lacs at the high redshifts from our samples, we estimate  to obtain 30-35 more high-$z$ BL Lacs from the 3LAC catalog which has $\sim$ 300 BL Lacs with no redshifts measurements. This will provide a total number of $\sim$ 50 high-$z$ BL Lacs, leading to the completion of this work and more importantly, will provide a large enough sample of BL Lacs at high redshifts for EBL studies.
\acknowledgements{AK and MA acknowledge funding under NASA contract 80NSSC17K0310. AD thanks the support of the Juan de la Cierva program from the Spanish MEC. DH, AK, MJ and VP thank the SARA consortium for their support providing required observational time.} 

\begin{deluxetable*}{llcccrc}[ht]
	\tablecolumns{7}
	\tablecaption{\label{tab:data} $\emph{Swift}$-UVOT and SARA Observations}
	\tablewidth{0pt}
	\tabletypesize{\small}
	\setlength{\tabcolsep}{0.07in} 
	\tablehead{
		\colhead{3FGL }   & \colhead{$\emph{$\emph{Swift}$}$} & \colhead{RA J2000} &	\colhead{Dec J2000} &	\colhead{$\emph{Swift}$ Date$^{a}$} & 	\colhead{SARA Date$^{a}$} & 	\colhead{$A_{V}$} \\
		\colhead{ (Name)}   & \colhead{(Name)} & \colhead{(hh:mm:ss)} &	\colhead{($^\circ: ^\prime: ''$)} &	\colhead{(UT)} & 	\colhead{(UT)} & 	\colhead{(mag)}
		}
	\startdata
      J0305.2$-$1607 & PKS 0302$-$16          & 03:05:15.04 & $-$16:08:16.3 & 2011$-$01$-$07.96 & 2016$-$12$-$04.2 & 0.12\\
      J0703.4$-$3914 & NVSS J070312$-$391418    & 07:03:12.66 & $-$39:14:18.9 & 2011$-$12$-$06.64 & 2016$-$12$-$01.1 & 0.35\\
      J0855.2$-$0718 & 3C 209                 & 08:55:58.31 & $-$07:26:36.9 & 2011$-$10$-$26.93 & 2016$-$12$-$01.2 & 0.09\\
      J0947.1$-$2542 & 1RXS J094709.2$-$254056   & 09:47:09.50 & $-$25:41:00.0 & 2013$-$10$-$06.80 & 2016$-$12$-$04.3 & 0.20\\
      J1155.4$-$3417 & 2FHL J1155.5$-$3417      & 11:55:20.47 & $-$34:17:19.9 & 2016$-$07$-$26.80 & 2017$-$02$-$18.1 & 0.22\\
      J1156.7$-$2250 & 1WHSP J115633.2$-$225004 & 11:56:33.20 & $-$22:50:04.5 & 2016$-$11$-$24.15 & 2017$-$02$-$18.2 & 0.13\\
      J1218.8$-$4827 & PMN J1219$-$4826         & 12:19:02.270 & $-$48:26:28.1 & 2013$-$06$-$26.88 & 2017$-$01$-$25.2 & 0.33\\
      J1518.0$-$2732 & TXS 1515$-$273           & 15:18:3.59 & $-$27:31:30.6 & 2014$-$09$-$30.87 & 2017$-$03$-$21.2 & 0.22\\
      J1723.7$-$7713 & TAN 1716$-$771           & 17:23:50.81  & $-$77:33:50.5 & 2011$-$05$-$13.81 & 2017$-$03$-$21.2 & 0.70\\
      J1759.1$-$4822 & PMN J1758$-$4820         & 17:58:58.45 & $-$48:21:12.4 & 2013$-$08$-$17.70 & 2017$-$03$-$21.3 & 0.55\\
      J1841.2+2910 & 2WHSP J184121.7+290940 & 18:41:21.70 & +29:09:41.0 & 2015$-$12$-$10.86 & 2017$-$07$-$14.0 & 0.64\\
      J1911.4$-$1908 & PMN J1911$-$1908         & 19:11:29.73 & $-$19:08:24.5 & 2014$-$07$-$30.45 & 2017$-$06$-$09.2 & 0.43\\
      J1955.0$-$1605 & 1RXS J195500.6$-$160328   & 19:55:00.58 & $-$16:03:37.9 & 2014$-$08$-$06.98 & 2017$-$06$-$09.3 & 0.56\\
      J2031.0+1937 & RX J2030.8+1935        & 20:30:57.13 & +19:36:1 & 2014$-$06$-$01.54 & 2017$-$07$-$14.1 & 0.25\\
      J2336.5$-$7620 & PMN J2336$-$7620         & 23:36:27.59 & $-$76:20:37.8 & 2014$-$07$-$19.39 & 2017$-$07$-$10.4 & 0.19 \\
      \enddata
      	\tablenotetext{a}{The observation dates for for $Swift$ and SARA correspond to the beginning of the exposures.}
    \end{deluxetable*}
    \clearpage
\begin{deluxetable*}{llllllrrrrr}[!ht]
	\tablecolumns{11}
	\tablecaption{\label{tab:mags} $\emph{Swift}$-UVOT and SARA photometry}
	\tablewidth{0pt}
	\tabletypesize{\footnotesize}
	\setlength{\tabcolsep}{0.03in} 
	\tablehead{
\colhead{3FGL Name}   & \colhead{$g^\prime$} & \colhead{$r^\prime$} &	\colhead{$i^\prime$} &	\colhead{$z^\prime$} & 	\colhead{$uvw2$} & 	\colhead{$uvm2$} & 	\colhead{$uvw1$} & 	\colhead{$u$} & \colhead{$b$} & \colhead{$v$}}
	\startdata
J0305.2$-$1607 & $>$20.89 & 18.96$\pm$ 0.06 & 18.40$\pm$ 0.06 & 18.68$\pm$ 0.17 & 22.24$\pm$ 0.12 & 21.99 $\pm$ 0.14 & 21.98$\pm$ 0.16 & 21.61$\pm$ 0.19 & 21.29$\pm$ 0.26 & $>$ 20.84\\
J0703.4$-$3914 & 18.71$\pm$ 0.03 & 17.54$\pm$ 0.01 & 15.76$\pm$ 0.02 & 16.38$\pm$ 0.03 & 20.42$\pm$ 0.12 & 20.46 $\pm$ 0.18 & 20.16$\pm$ 0.17 & 19.31$\pm$ 0.12 & 18.93$\pm$ 0.14 & 18.61$\pm$ 0.19\\
J0855.2$-$0718 & 19.87$\pm$ 0.12 & 18.93$\pm$ 0.05 & 18.34$\pm$ 0.08 & 18.86$\pm$ 0.28 & 22.09$\pm$ 0.22 & 21.37 $\pm$ 0.24 & 20.86$\pm$ 0.17 & 20.40$\pm$ 0.19 & 20.04$\pm$ 0.25 & $>$ 19.61\\
J0947.1$-$2542 & 16.74$\pm$ 0.02 & 16.50$\pm$ 0.01 & 17.41$\pm$ 0.01 & 16.49$\pm$ 0.03 & 17.73$\pm$ 0.06 & 17.75 $\pm$ 0.08 & 17.45$\pm$ 0.07 & 17.09$\pm$ 0.06 & 16.77$\pm$ 0.07 & 16.78$\pm$ 0.12\\
J1155.4$-$3417 & 18.19$\pm$ 0.02 & 17.95$\pm$ 0.01 & 17.84$\pm$ 0.02 & 17.72$\pm$ 0.04 & 19.80$\pm$ 0.14 & 19.64 $\pm$ 0.15 & 19.36$\pm$ 0.17 & 18.91$\pm$ 0.18 & 18.22$\pm$ 0.17 & 18.03$\pm$ 0.24\\
J1156.7$-$2250 & 19.17$\pm$ 0.04 & 18.91$\pm$ 0.02 & 18.76$\pm$ 0.03 & 18.73$\pm$ 0.07 & 20.77$\pm$ 0.16 & 20.37 $\pm$ 0.17 & 20.30$\pm$ 0.20 & 19.64$\pm$ 0.19 & 19.20$\pm$ 0.36& $>$ 18.91\\
J1218.8$-$4827 & 17.24$\pm$ 0.04 & 16.46$\pm$ 0.02 & 16.44$\pm$ 0.02 & 16.10$\pm$ 0.04 & 18.58$\pm$ 0.09 & 18.56 $\pm$ 0.18 & 18.17$\pm$ 0.10 & 18.01$\pm$ 0.12 & 17.38$\pm$ 0.11 & 17.39$\pm$ 0.18\\
J1518.0$-$2732 & 17.44$\pm$ 0.12 & 16.51$\pm$ 0.03 & 16.20$\pm$ 0.04 & 15.97$\pm$ 0.07 & 18.67$\pm$ 0.09 & 18.75 $\pm$ 0.11 & 18.46$\pm$ 0.12 & 17.81$\pm$ 0.10 & 17.60$\pm$ 0.14 & 16.97$\pm$ 0.14\\
J1723.7$-$7713 & 18.42$\pm$ 0.05 & 17.99$\pm$ 0.02 & 17.92$\pm$ 0.03 & 17.53$\pm$ 0.06 & 19.35$\pm$ 0.22 & $>$ 19.79  & 19.34$\pm$ 0.28 & $>$ 19.33 & 18.49$\pm$ 0.29 & $>$ 17.86\\
J1759.1$-$4822 & 15.57$\pm$ 0.04 &  13.03 $\pm$ 0.02&  15.24 $\pm$ 0.03 & 15.26$\pm$ 0.06 & $>$ 17.87 & $>$ 17.52 & $>$ 17.33 & $>$ 16.97 & 16.17$\pm$ 0.31 & $>$ 15.65\\
J1841.2+2910 & 16.83$\pm$ 0.01 & 17.09$\pm$ 0.01 & 17.10$\pm$ 0.01 & 16.90$\pm$ 0.03 & 18.14$\pm$ 0.18 & $>$18.57  & 17.57$\pm$ 0.18 & 17.37$\pm$ 0.20 & 16.77$\pm$ 0.19 & $>$16.90\\
J1911.4$-$1908 & 18.24$\pm$ 0.03 & 17.62$\pm$ 0.01 & 17.30$\pm$ 0.01 & 16.97$\pm$ 0.02 & 19.83$\pm$ 0.27 & 19.73 $\pm$ 0.26 & $>$19.46 & 18.69$\pm$ 0.27 & 18.34$\pm$ 0.29 & 17.66$\pm$ 0.36\\
J1955.0$-$1605 & 17.72$\pm$ 0.05 & 17.35$\pm$ 0.02 & 17.04$\pm$ 0.02 & 16.69$\pm$ 0.03 & 19.02$\pm$ 0.20 & $>$19.22  & 18.35$\pm$ 0.26 & 18.17$\pm$ 0.20 & 17.77$\pm$ 0.24 & $>$ 17.58\\
J2031.0+1937 & 17.52$\pm$ 0.01 & 17.93$\pm$ 0.01 & 17.75$\pm$ 0.01 & 17.57$\pm$ 0.04 & 18.33$\pm$ 0.13 & 18.45 $\pm$ 0.24 & 18.17$\pm$ 0.19 & 17.50$\pm$ 0.17 & 17.45$\pm$ 0.30 & $>$16.95\\
J2336.5$-$7620 & 18.06$\pm$ 0.05 & 16.53$\pm$ 0.02 & 17.34$\pm$ 0.03 & 17.24$\pm$ 0.08 & 19.74$\pm$ 0.16 & 19.43 $\pm$ 0.22 & 19.12$\pm$ 0.17 & 18.82$\pm$ 0.18 & 18.35$\pm$ 0.20 & 18.18$\pm$ 0.32\\
\enddata
\end{deluxetable*}

\begin{deluxetable*}{ll|ccll|lclr}
	\tablecolumns{10}
	\tabletypesize{\small}
	\tablewidth{0pt}
	\tablecaption{\label{tab:result}SED fitting}
	\setlength{\tabcolsep}{0.05in} 
	\tablehead{
		\colhead{3FGL Name}  & \colhead{$z_{\rm phot, best}^a$}   & \multicolumn{4}{c}{Power Law Template} & \multicolumn{4}{c}{Galaxy Template} \\
		\colhead{} & \colhead{} & \colhead{$z_{\rm phot}$ $^b$} & \colhead{$\chi^2$} & \colhead{P$_{\rm z}^c$} & \colhead{$\beta^d$} & \colhead{$z_{\rm phot}$ $^b$} & \colhead{$\chi^2$} & \colhead{P$_{\rm z}$  $^c$} & \colhead{model}} 
	\startdata
	\multicolumn{10}{c}{\textbf{\small Sources with confirmed photometric redshifts}} \\
	\hline
	J1155.4$-$3417     &  ${1.83}^{+0.10}_{-0.13}$ & ${1.83}^{+0.10}_{-0.13}$ &   20.0 &  99.9 &  0.65 & ${1.50}^{+0.02}_{-0.03}$ & 28.8  & 100.0  & pl\_QSOH\_template\_norm.sed   \\ 
		J1156.7$-$2250     & ${1.73}^{+0.11}_{-0.19}$  & ${1.73}^{+0.11}_{-0.19}$ &   9.2 &  98.5 &  0.70 & ${1.50}^{+0.04}_{-0.07}$ & 7.1  & 99.9  & pl\_QSOH\_template\_norm.sed   \\ 
	\hline \hline\multicolumn{10}{c}{\textbf{\small Sources with photometric redshifts upper limits}} \\
	\hline
J0305.2$-$1607 & \nodata  & ${0.85}^{+ 0.01}_{- 0.18} $ &  565.2 &  45.8 & 0.80 &  ${1.22 }^{+0.03 }_{-0.02 }$ & 297.3& 93.6 & S0\_template\_norm.sed   \\
J0703.4$-$3914 & \nodata  & ${4.00}^{+ 0.00}_{- 0.01} $ & 267.6 & 100.0 & 0.45 & ${0.61 }^{+0.04 }_{- 0.05} $ & 25.4& 89.5 & Sey2\_template\_norm.sed  \\
J0855.2$-$0718 & \nodata  & ${3.86}^{+ 0.00}_{- 0.00} $ &  128.0 &  82.3 & 1.10 & ${0.45 }^{+0.05 }_{-0.01 } $   & 39.3 & 99.9 & M82\_template\_norm.sed\\
J0947.1$-$2542 & \nodata& ${1.63}^{+ 0.10}_{- 0.08} $ & 70.6 &  99.9 & 0.45  & ${1.33 }^{+0.05 }_{-0.04 } $  & 24.9 & 63.6 & pl\_TQSO1\_template\_norm.sed \\
J1218.8$-$4827 & \nodata & ${1.23}^{+ 0.21}_{- 0.60} $ &  181.7 &  50.9 & 1.40 & ${0.26 }^{+0.01 }_{-0.01 } $  & 28.8 & 100.0 & pl\_QSOH\_template\_norm.sed\\
J1518.0$-$2732 & $<$1.1 & ${0.05}^{+ 1.03}_{- 0.05} $ &   22.9 &  16.8 & 1.85 & ${0.19 }^{+ 0.10}_{-0.06} $  & 11.07 & 98.0& Mrk231\_template\_norm.sed \\
J1723.7$-$7713 & \nodata & ${2.74}^{+ 0.21}_{- 0.04} $ &   44.3 &  99.5 & 0.80 & ${0.01}^{+0.03 }_{-0.01 } $ & 19.8 & 99.8  & Spi4\_template\_norm.sed \\
J1759.1$-$4822 & \nodata & ${3.86}^{+ 0.00}_{- 0.00} $   &   \nodata &  \nodata & \nodata & \nodata & \nodata & \nodata&\nodata  \\
J1841.2+2910 & \nodata & ${2.06}^{+ 0.08}_{- 0.10} $ &  176.7 & 100.0 & \nodata & ${0.45 }^{+0.03 }_{-0.01 } $  & 244.9 & 100.0 & S0\_80\_QSO2\_20.sed \\
J1911.4$-$1908 & $<$1.6  & ${1.33}^{+ 0.28}_{- 1.33} $ &   21.2 &  41.0 & 1.60 & ${ 0.06}^{+0.03 }_{-0.02} $& 7.1 & 100.0 &  Mrk231\_template\_norm.sed  \\
J1955.0$-$1605 & $<$1.2  & ${0.03}^{+ 1.13}_{- 0.03} $ &   25.7 &  13.6 & 1.65 & ${1.26 }^{+0.04 }_{-0.03 } $ & 36.2 & 100.0 &  I22491\_90\_TQS01\_10.sed\\
J2031.0+1937 & \nodata & ${2.03}^{+ 0.10}_{- 0.08} $ &   247.8 &  100.0 & \nodata & ${1.93 }^{+0.02 }_{-0.05:w } $  & 255.9 & 100.0 & I22491\_70\_TQS01\_20.sed\\
J2336.5$-$7620 & $<$1.5 & ${1.33}^{+0.22 }_{-0.11 } $ & 9.5   & 60.8   & 1.35 & ${0.08 }^{+0.05 }_{-0.04 } $   & 8.9 & 100.0 &  Spi4\_template\_norm.sed\\
  \enddata
	\tablenotetext{a}{Best photometric redshift.}
	\tablenotetext{b}{Photometric redshifts with 2$\sigma$ confidence level}
	\tablenotetext{c}{Redshift probability density at $z_{phot} \pm 0.1(1+z_{phot})$}
	\tablenotetext{d}{Spectral slope for power law model of the form $F_{\lambda} \propto \lambda^{-\beta}$}
\end{deluxetable*}
\clearpage
\bibliographystyle{apj}
\bibliography{bibliography}
\end{document}